\documentclass[12pt]{iopart}

\usepackage{graphicx}

\begin{document}
\title[Cosmic expansion from boson and fermion fields]
{Cosmic expansion from boson and fermion fields}

\author{Rudinei C. de Souza\dag\footnote{rudijantsch@gmail.com}\ and Gilberto M. Kremer\dag
\footnote{kremer@fisica.ufpr.br}
}

\address{\dag\ Departamento de F\'{\i}sica, Universidade Federal do Paran\'a,
 Curitiba, Brazil}

\begin{abstract}
This paper consists in analyzing an action that describes boson and fermion fields minimally coupled to the gravity and a common matter field. The self-interaction potentials of the fields are not chosen \emph{a priori} but from the Noether symmetry approach. The Noether forms of the potentials allow the boson field to play the role of dark energy and matter and the fermion field to behave as standard matter. The constant of motion and the cyclic variable associated with the Noether symmetry allow the complete integration of the field equations, whose solution produces a Universe with alternated periods of accelerated and decelerated expansion.
\end{abstract}

 \date{\today}



\maketitle

\section{Introduction}

In the present day the cosmology faces two important problems. The first is that the common matter cannot explain the galaxy rotation curves and so another type of matter is needed \cite{1}. Although this is an old problem \cite{2}, the nature of this another type of matter is unknown until the present and several candidate particles are proposed in the literature \cite{2.1}. Nowadays the most accepted explanation that cosmologists have for the galaxy rotation curves is that there exists a strange matter field that interacts only gravitationally with the known matter -- the so-called dark matter. Some alternative theories to the dark matter fluid are the Modified Newtonian Dynamics (MOND) \cite{2.2}, the $f(R)$ \cite{2.3} and $f_i(R)$ \cite{2.4} (non-minimally coupled matter) theories which consider the dark matter as a geometrical effect. The recent data from the measures of the matter contained in the galaxies through the gravitational lensing effect strongly support the existence of the dark matter \cite{3,4}.

The second problem is that the astronomical observations show that the expansion of the Universe is accelerated in the late time \cite{5,6}. The standard cosmology cannot explain what causes this acceleration and cosmologists have to look for a satisfactory explanation for this observed fact. The first attempt of explaining it came from the revival of the cosmological constant. Unfortunately, such an idea soon presented inconsistences \cite{2.a} and new models appeared in the literature, but no one has proved to be the definitive one yet \cite{2.b}. Among these models, a well known is the quintessence model \cite{7,8}, which can present several acceptable forms for the potentials, proposed phenomenologically or inspired by fundamental theories \cite{8.1,8.2,8.3}. The models with fermion fields \cite{7.3,7.3.1,7.3.2}, tachyon fields \cite{7.4,7.4.1} (k-essence), Chaplygin gas \cite{7.6,7.6.1} and van der Waals gas \cite{7.a,7.b} are also alternatives to the cosmological constant. All these models essentially try to model an exotic component with negative pressure which composes the most part of the energy density of the Universe -- the so-called dark energy. The $f(R)$ theories \cite{7.7,7.7.1} also propose an explanation of geometric origin for the accelerated expansion.

We starts this work from an action with boson (quintessence-type) and Dirac fermion fields minimally coupled to the gravity and a common matter field. One considers a spatially flat, homogeneous and isotropic Universe (in according to the observations) and a point-like Lagrangian is obtained from the action. The undefined potentials in the Lagrangian are selected from the existence condition for the Noether symmetry -- interesting results using this approach can be found in the references \cite{7.4.1,9,9.1,11,12,10,13.1}. The selected potentials indicate that the boson field can describe the dark energy and an additional matter field and the fermion field behaves as the standard matter. From this result we describe the dark sector by bosons and fermions, with the bosons composing a dark energy--matter field and the fermions playing the role of dark matter. This type of fermionic dark matter (but with additional interactions) is also studied in \cite{14}. The standard matter behavior of Dirac fermion fields can be seen in other contexts \cite{12,12.1,12.2}.

Since the potentials satisfying the Noether condition imply that the Lagrangian presents a Noether symmetry, there is a constant of motion and a cyclic variable can be obtained through a coordinate transformation in the configuration space. This can help us integrate the dynamical system. Then, using these additional informations furnished by the Noether symmetry, the field equations of our original problem are completely integrated. The result is that the Universe described by the proposed action -- satisfying such a symmetry -- evolves with alternated periods of accelerated and decelerated expansion.

The organization of the paper is the following: in the second section the action of the model is presented and the field equations from a point-like Lagrangian for a flat Friedmann-Robertson-Walker (FRW) metric are derived. The selection of the potentials from the Noether symmetry condition is performed in the third section and the identification of the fermion field with the dark matter is discussed in the fourth section. In the fifth section the field equations are integrated via constant of motion and cyclic variable and the analysis of the corresponding solution is fulfilled in the sixth section. The paper is closed with the conclusions in the seventh section.

In this work we adopt the signature $(+, -, -, -)$ and the natural units $8 \pi G=c=\hbar=1$.

\section{Action and field equations}

Let us start from a general action describing a boson field and a fermion field minimally coupled to the gravity

\begin{eqnarray}\nonumber
S=&\int d^4x\sqrt{-g}\ \Bigg\{\frac{R}{2}+\frac{1}{2}g^{\mu\nu}\partial_\mu\phi\partial_\nu\phi-U(\phi)\\
&+\frac{\imath}{2}\left[\overline\psi\Gamma^\mu D_\mu\psi-(\overline D_\mu\overline\psi)\Gamma^\mu\psi\right]-V(\Psi)+\mathcal{L}_M\Bigg\}, \label{ga}
\end{eqnarray}
where $\phi$ represents a boson field, $\psi$ and $\overline\psi=\psi^\dag\gamma^0$ denote the spinor field and its adjoint, respectively, which represent a Dirac fermion field, and  $\mathcal{L}_M$ describes a common matter field. $R$ is the Ricci scalar, the dagger denotes complex conjugation transpose and $\Psi=\overline\psi \psi$. Further, $U$ and $V$ are the self-interaction potentials of the boson and fermion fields, respectively. In (\ref{ga}) the covariant derivatives $D_\mu$ and $\overline D_\mu$ are defined by
\begin{equation}
D_\mu\psi=\left(\partial_\mu-\Omega_\mu\right)\psi, \qquad
\overline D_\mu\overline\psi=\partial_\mu\overline\psi+\overline\psi\Omega_\mu,
\end{equation}
where $\Omega_\mu$ is the spin connection
\begin{equation}
\Omega_\mu=-\frac{1}{4}g_{\sigma\nu}\left[\ \Gamma^\nu_{\mu\lambda}-e^\nu_b(\partial_\mu e^b_\lambda)\right ]\Gamma^\sigma\Gamma^\lambda,
\end{equation}
with  $e^a_\mu$ denoting the "vierbein" and $\Gamma^\mu=e_a^\mu\gamma^a$. The $\gamma^a$ are the Pauli-Dirac matrices.

By varying the action (\ref{ga}) with respect to the metric tensor $g_{\mu\nu}$ one obtains the Einstein's field equations
\begin{equation}
R_{\mu\nu}-\frac{1}{2}g_{\mu\nu}R=-T_{\mu\nu}, \label{efe}
\end{equation}
with $T_{\mu\nu}=T_{\mu\nu}^M+T_{\mu\nu}^\psi+T_{\mu\nu}^\phi$, where the letters $M, \psi$ and $\phi$ label the energy-momentum tensors of the common matter, boson and fermion fields, respectively, which are defined by

\begin{equation}
T_{\mu\nu}^M=\frac{2}{\sqrt{-g}}\frac{\delta\left(\sqrt{-g}\mathcal{L}_M\right)}{\delta
g^{\mu\nu}}, \label{emt1}
\end{equation}

\begin{equation}
T_{\mu\nu}^{\psi}=\frac{\imath}{4}\left[ \ \overline\psi\Gamma_\nu D_\mu\psi+\overline\psi\Gamma_\mu D_\nu\psi-\overline
D_\mu \overline\psi\Gamma_\nu\psi-\overline D_\nu \overline\psi\Gamma_\mu\psi \ \right]-g_{\mu\nu}\mathcal{L}_\psi, \label{emt2}
\end{equation}

\begin{equation}
T_{\mu\nu}^\phi=\partial_\mu\phi\partial_\nu\phi-\left(\frac{1}{2}\partial_\theta\phi\partial^\theta\phi-
U\right)g_{\mu\nu}, \label{emt3}
\end{equation}
where $\mathcal{L}_\psi$ represents the Dirac Lagrangian
\begin{equation}
\mathcal{L}_\psi=\frac{\imath}{2}\left[\ \overline\psi\Gamma^\mu D_\mu\psi-(\overline D_\mu\overline\psi)\Gamma^\mu\psi \ \right]-V.
\end{equation}

From the action (\ref{ga}), for a flat FRW metric, with the fields $\phi$ and $\psi$ spatially homogeneous and a pressureless common matter field, we can write a point-like Lagrangian of the form

\begin{equation}
\mathcal{L}=3a\dot{a}^2-a^3\left(\frac{\dot\phi^2}{2}-U\right)+\frac{\imath}{2}a^3\left(
\dot{\overline\psi}\gamma^0 \psi -
\overline\psi\gamma^0\dot{\psi}\right)+a^3V+\rho_M^0, \label{plL}
\end{equation}
which furnishes the same dynamical equations that those from the Einstein's field equations (\ref{efe}). The point represents time derivatives and $a$ is the scale factor. Here $\rho_M^0$ is the energy density of the common matter at an initial instant (this comes from the common matter energy density: $\rho_M=\rho_M^0/a^3$).

Imposing that the "energy function" associated with the Lagrangian (\ref{plL}) is null,
the result is the Friedmann's equation, i.e.,
\begin{eqnarray}\nonumber
E_\mathcal{L}=\frac{\partial \mathcal{L}}{\partial \dot{a}}
\dot{a}+\frac{\partial \mathcal{L}}{\partial \dot{\phi}}
\dot{\phi}+\dot{\overline\psi} \frac{\partial
\mathcal{L}}{\partial\dot{\overline\psi}}+\frac{\partial
\mathcal{L}}{\partial\dot{\psi}}\dot{\psi}-\mathcal{L}=0,
\\ \Longrightarrow \left(\frac{\dot a}{a}\right)^2=\frac{\rho_M+\rho_\psi+\rho_\phi}{3},\label{fride}
\end{eqnarray}
with $\rho_\psi$ and $\rho_\phi$ denoting the energy densities, defined as

\begin{equation}
\rho_\psi=V, \qquad \rho_\phi=\frac{1}{2}\dot\phi^2+U. \label{ed}
\end{equation}
In according to the energy-momentum tensors (\ref{emt2}) and (\ref{emt3}).

From the Euler-Lagrange equation for $a$, applied to (\ref{plL}), one has the acceleration equation

\begin{equation}
\frac{\ddot{a}}{a}=-\frac{\rho_M+\rho_\psi+\rho_\phi+3\left(p_\psi+p_\phi\right)}{6}, \label{acce}
\end{equation}
and for $\overline\psi$ and $\psi$, we have the Dirac's equations
for the spinor field and its adjoint coupled to the gravitational field, respectively,
\begin{eqnarray}
&\dot{\psi}+\frac{3}{2}\left(\frac{\dot a}{a}\right)\psi+\imath\gamma^0\psi\frac{dV}{d\Psi}=0,\\
&\dot{\overline\psi}+\frac{3}{2}\left(\frac{\dot a}{a}\right)\overline\psi-\imath\overline\psi\gamma^0\frac{dV}{d\Psi}=0.
\end{eqnarray}
In (\ref{acce}) the pressures are defined as

\begin{equation}
p_\psi=\Psi\frac{dV}{d\Psi}-V, \qquad p_\phi=\frac{\dot\phi^2}{2}-U. \label{press}
\end{equation}
Also in according to (\ref{emt2}) and (\ref{emt3}).

Finally, the Euler-Lagrange equation for $\phi$ gives

\begin{equation}
\ddot\phi+3\left(\frac{\dot a}{a}\right)\dot\phi+\frac{dU}{d\phi}=0,\label{KGe}
\end{equation}
which is just the Klein-Gordon equation for the field $\phi$.

\section{Selection of potentials}

One can express the Lagrangian (\ref{plL}) in terms of the components of the spinor field $\psi=(\psi_1,
\psi_2, \psi_3, \psi_4)^{T}$ and its adjoint $\overline
\psi=(\psi_1^*, \psi_2^*, -\psi_3^*,
-\psi_4^*)$. So it takes the form

\begin{equation}
\mathcal{L}=3a\dot{a}^2-a^3\left(\frac{\dot\phi^2}{2}-U\right)+\frac{\imath}{2}a^3\sum_{i=1}^{4}\left(\dot{\psi_i^*}\psi_i-\psi_i^*
\dot{\psi_i}\right)+a^3V+\rho_M^0. \label{plL1}
\end{equation}
Thus the dynamics of the system is now described by 10 coordinates, i.e., the configuration space of the system is represented by $\left\{a, \phi, \psi_j^*\ , \psi_j\right\}$, with $j=1,2,3,4$.

The undefined forms of the potentials can be selected by the requirement of the existence of a
Noether symmetry for a given Lagrangian. This approach constrains the possible
functions for the potentials at the same time that it provides a conserved quantity
associated with the selected forms. Since the dynamical system presents a Noether symmetry, one has a constant of motion and a cyclic variable is allowed, which can help us integrate the system. A Noether symmetry exists for a given Lagrangian of the form $\mathcal{L}=\mathcal{L}(q_k, \dot q_k)$ if the condition $L_\textbf{x}\mathcal{L}=0$ is satisfied, with $L_\textbf{x}$ designating the Lie derivative with respect to
the vector field $\textbf{X}$ (called infinitesimal generator of symmetry) defined by
\begin{equation}
\textbf{X}=\sum_k\left(\alpha_k\frac{\partial}{\partial q_k}+
\frac{d\alpha_k}{dt}\frac{\partial}{\partial\dot q_k}\right),
\end{equation}
where the $\alpha_k$'s are functions of the generalized coordinates
$q_k$.

From the E-L equations one arrives at the relation

\begin{equation}
\frac{d}{dt}\left(\sum_k\alpha_k\frac{\partial\mathcal{L}}{\partial \dot q_k}\right)=L_\textbf{x}\mathcal{L}=0,
\end{equation}
and the conserved quantity (or constant of motion) associated with the Noether symmetry is obtained
\begin{equation}
M_0=\sum_k\alpha_k\frac{\partial\mathcal{L}}{\partial \dot q_k}. \label{constm}
\end{equation}

The new set of variables $\left\{Q_k(q_l)\right\}$ for the configuration space, such that one of the variables is cyclic, obeys the following system of differential equations

\begin{equation}
\sum_l\left(\alpha_l\frac{\partial u_{k-1}}{\partial q_l}\right)=0,\nonumber\\
\sum_l\left(\alpha_l\frac{\partial z}{\partial q_l}\right)=1, \label{cyclicv}
\end{equation}
where $z$ is the cyclic variable. The range of the index $l$ is the same of that of $k$ and the $u_{k-1}$ and $z$ comprehend the new coordinates $Q_k(q_l)$ of the configuration space.

Starting from the Noether symmetry condition
$L_{\textbf{x}}\mathcal{L}=\textbf{X}\mathcal{L}=0$ applied to the
Lagrangian (\ref{plL1}), with the infinitesimal generator of symmetry
$\textbf{X}$ defined for our problem as

\begin{eqnarray}\nonumber
\textbf{X}=&\ C_0\frac{\partial}{\partial
a}+\dot{C_0}\frac{\partial}{\partial
\dot{a}}+D_0\frac{\partial}{\partial
\phi}+\dot{D_0}\frac{\partial}{\partial
\dot{\phi}}\\
&+\sum_{i=1}^{4}\left(C_i\frac{\partial}{\partial
\psi_i^*}+D_i\frac{\partial}{\partial
\psi_i}+\dot{C_i}\frac{\partial}{\partial
\dot{\psi_i^*}}+\dot{D_i}\frac{\partial}{\partial
\dot{\psi_i}}\right),
\end{eqnarray}
we have the following system of partial differential equations

\begin{equation}
C_0+2a\frac{\partial C_0}{\partial a}=0, \qquad 3C_0+2a\frac{\partial D_0}{\partial \phi}=0, \qquad 6\frac{\partial C_0}{\partial \phi}-a^2\frac{\partial D_0}{\partial a}=0, \label{sc1}
\end{equation}

\begin{equation}
\sum_{i=1}^4\left({\partial C_i\over \partial a}\psi_i- {\partial
D_i\over \partial a}\psi_i^*\right)=0, \qquad \sum_{i=1}^4\left({\partial C_i\over \partial \phi}\psi_i- {\partial
D_i\over \partial \phi}\psi_i^*\right)=0, \label{sc2}
\end{equation}

\begin{equation}
3C_0\psi_j+aD_j+a\sum_{i=1}^4\left({\partial C_i\over \partial
\psi_j^*}\psi_i- {\partial D_i\over \partial
\psi_j^*}\psi_i^*\right)=0, \label{sc3}
\end{equation}

\begin{equation}
3C_0\psi_j^*+aC_j-a\sum_{i=1}^4\left({\partial C_i\over
\partial \psi_j}\psi_i- {\partial D_i\over \partial
\psi_j}\psi_i^*\right)=0, \label{sc4}
\end{equation}

\begin{equation}
\frac{\partial C_0}{\partial \psi_j^*}=0, \qquad \frac{\partial C_0}{\partial \psi_j}=0,\qquad \frac{\partial D_0}{\partial \psi_j^*}=0, \qquad \frac{\partial D_0}{\partial \psi_j}=0, \label{sc5}
\end{equation}

\begin{equation}
3C_0(U+V)+aD_0\frac{dU}{d\phi}+a\sum_{i=1}^4\left(C_i\epsilon_i\psi_i+D_i\epsilon_i\psi_i^*\right)\frac{dV}{d\Psi}=0, \label{sc6}
\end{equation}
where the symbol $\epsilon_i$ takes the values

\begin{eqnarray}\nonumber
\epsilon_i=+1\qquad\hbox{for}\qquad i=1,2;\\
\epsilon_i=-1\qquad\hbox{for}\qquad i=3,4.
\end{eqnarray}

 One can see from equations (\ref{sc5}) that the coefficients $C_0$ and $D_0$ are functions only of $a$ and $\phi$. Then, assuming that $C_0$ and $D_0$ are separable functions

\begin{equation}
C_0=c_1(a)c_2(\phi), \qquad D_0=d_1(a)d_2(\phi),
\end{equation}
we obtain the solution for the system (\ref{sc1})-(\ref{sc6})

\begin{equation}
C_0=\frac{Ae^{\alpha\phi}+Be^{-\alpha\phi}}{\sqrt{6a}},
\end{equation}

\begin{equation}
D_0=-\frac{Ae^{\alpha\phi}-Be^{-\alpha\phi}}{a^{3/2}},
\end{equation}

\begin{equation}
C_j=-\alpha\frac{Ae^{\alpha\phi}+Be^{-\alpha\phi}}{a^{3/2}}{\psi_j^* }+\beta\epsilon_j\psi_j^*,
\end{equation}

\begin{equation}
D_j=-\alpha\frac{Ae^{\alpha\phi}+Be^{-\alpha\phi}}{a^{3/2}}{\psi_j}-\beta\epsilon_j\psi_j,
\end{equation}

\begin{equation}
U=U_0\left(Ae^{\alpha\phi}-Be^{-\alpha\phi}\right)^2, \qquad V=V_0\Psi, \label{potentials}
\end{equation}
with $A, B, U_0, V_0$ and $\beta$ being constants and $\alpha=\frac{1}{2}\sqrt{\frac{3}{2}}$.

These are the coefficients and potentials that satisfy the Noether symmetry condition. Now, if the potentials (\ref{potentials}) are taken for the Lagrangian (\ref{plL1}), it presents a Noether symmetry.

\section{Dark matter as a fermion field}

The solution for the potential of the fermion field is $V=V_0\Psi$, which is essentially a term of mass. Thus we can replace the constant $V_0$ for $m$ in order to identify this constant with the mass of the fermions. Once it is done, we determine from equations $(\ref{ed})_1$ and $(\ref{press})_1$ that the fermion field has its energy density and pressure given by

\begin{equation}
\rho_\psi=m\Psi, \qquad p_\psi=0,
\end{equation}
characterizing a field of pressureless matter. Note that this form for $\rho_\psi$ requires that $\Psi$ must be a positive quantity.

Observe that we did not assume \emph{a priori} that the fermion field is a massive field, whereas the potential takes this information. This characteristic of the field comes exclusively from the symmetry condition when it constrains the potential to a term of mass.

This result \emph{suggests} that the fermion field behaves as a standard matter field. But obviously this field has a nature different from that of the common matter since it is composed exclusively by fermionic particles. Thus, having in view that these fermions produce an additional pressureless matter, one may identify the fermion field with the dark matter. For this propose, we must assume that this field describes fermionic particles that interact only gravitationally with the common matter or have very weak non-gravitational interactions with it.

On the other hand, the field $\phi$ received from the Noether condition the potential $(\ref{potentials})_1$, which allows an accelerated expansion. Hence, from this point on, the dark sector will be identified with the fields $\phi$ and $\psi$.

\section{Searching for exact solutions}

Taking $U$ and $V$ given by (\ref{potentials}), we have to solve the following system of coupled differential equations

\begin{equation}
3\left(\frac{\dot a}{a}\right)^2=\frac{\rho_M^0}{a^3}+m\Psi+\frac{\dot\phi^2}{2}+U_0\left(Ae^{\alpha\phi}-Be^{-\alpha\phi}\right)^2, \label{fe}
\end{equation}

\begin{equation}
2\frac{\ddot{a}}{a}+\left(\frac{\dot a}{a}\right)^2+\frac{\dot\phi^2}{2}-U_0\left(Ae^{\alpha\phi}-Be^{-\alpha\phi}\right)^2=0, \label{ae}
\end{equation}

\begin{equation}
\ddot\phi+3\left(\frac{\dot a}{a}\right)\dot\phi+2\alpha U_0\left(A^2e^{2\alpha\phi}-B^2e^{-2\alpha\phi}\right)=0, \label{kge}
\end{equation}

\begin{equation}
\dot{\psi}+\frac{3}{2}\left(\frac{\dot a}{a}\right)\psi+\imath m\gamma^0\psi=0, \qquad \dot{\overline\psi}+\frac{3}{2}\left(\frac{\dot a}{a}\right)\overline\psi-\imath m\overline\psi\gamma^0=0. \label{dirac1,2}
\end{equation}

  Once with these forms for $U$ and $V$ the dynamical system presents a Noether symmetry, one has an additional dynamical equation provided by the constant of motion, namely
\begin{eqnarray}\nonumber
M_0&=C_0\frac{\partial\mathcal{L}}{\partial
\dot{a}}+D_0\frac{\partial\mathcal{L}}{\partial
\dot{\phi}}+\sum_{i=1}^{4}\left(C_i\frac{\partial\mathcal{L}}{\partial
\dot{\psi_i^*}}+D_i\frac{\partial\mathcal{L}}{\partial
\dot{\psi_i}}\right)\nonumber\\
&=\left(Ae^{\alpha\phi}+Be^{-\alpha\phi}\right)\sqrt{6a}\dot a+\left(Ae^{\alpha\phi}-Be^{-\alpha\phi}\right)a^{3/2}\dot\phi+\imath\beta\ a^3\Psi, \label{cm}
\end{eqnarray}
which is determined from equation (\ref{constm}). In this equation one can make $\beta=\imath\beta_0$, with $\beta_0$ being a real constant, in order to have a constant of motion whose value is real, whereas $\Psi$ is a real positive quantity.

  As it was exposed before, if the system presents a Noether symmetry, there is a cyclic variable that reduces the dynamics of the system, which is obtained from a transformation of variables. The transformation that takes the configuration space from $\{a, \phi, \psi_j^*\ , \psi_j\}$ to $\{z, u, v_j, w_j\}$ satisfies the system

\begin{equation}
C_0\frac{\partial u}{\partial
a}+D_0\frac{\partial u}{\partial
\phi}+\sum_{i=1}^{4}\left(C_i\frac{\partial u}{\partial
\dot{\psi_i^*}}+D_i\frac{\partial u}{\partial \dot{\psi_i}}\right)=0,\label{v1}
\end{equation}

\begin{equation}
C_0\frac{\partial v_j}{\partial
a}+D_0\frac{\partial v_j}{\partial
\phi}+\sum_{i=1}^{4}\left(C_i\frac{\partial v_j}{\partial
\dot{\psi_i^*}}+D_i\frac{\partial v_j}{\partial
\dot{\psi_i}}\right)=0,\label{v2}
\end{equation}

\begin{equation}
C_0\frac{\partial w_j}{\partial
a}+D_0\frac{\partial w_j}{\partial
\phi}+\sum_{i=1}^{4}\left(C_i\frac{\partial w_j}{\partial
\dot{\psi_i^*}}+D_i\frac{\partial w_j}{\partial
\dot{\psi_i}}\right)=0,\label{v3}
\end{equation}

\begin{equation}
C_0\frac{\partial z}{\partial
a}+D_0\frac{\partial z}{\partial
\phi}+\sum_{i=1}^{4}\left(C_i\frac{\partial z}{\partial
\dot{\psi_i^*}}+D_i\frac{\partial z}{\partial
\dot{\psi_i}}\right)=1,\label{v4}
\end{equation}
with $z$ being the cyclic variable. This system is obtained by applying (\ref{cyclicv}).

Using the constant of motion (\ref{cm}) and a transformation of variables satisfying (\ref{v1})-(\ref{v4}), we will look for a solution to the field equations (\ref{fe})-(\ref{dirac1,2}).

Firstly, the Dirac's equations $(\ref{dirac1,2})_1$ and $(\ref{dirac1,2})_2$ can reduced to an equation for $\Psi$

\begin{equation}
\dot{\Psi}+3\left(\frac{\dot a}{a}\right)\Psi=0,
\end{equation}
which is easily integrated in terms of $a$, giving

\begin{equation}
\Psi=\frac{\Psi_0}{a^3},
\end{equation}
where $\Psi_0$ is a positive constant. Or, by solving equation $(\ref{dirac1,2})_1$ in terms of $a$, the solution for $\psi$ is

\begin{equation}
 \psi=\left(%
\begin{array}{c}
\psi_{1}^0e^{-\imath mt}  \\
\psi_{2}^0e^{-\imath mt}  \\
\psi_{3}^0e^{\imath mt}  \\
\psi_{4}^0e^{\imath mt}  \\
\end{array}%
\right)a^{-3/2}\label{spinor1},
\end{equation}
with $\psi_j^0$ being constants that satisfy the relation $\Psi_0=\psi_1^{*0}\psi_1^0+\psi_2^{*0}\psi_2^0-\psi_3^{*0}\psi_3^0-\psi_4^{*0}\psi_4^0$.

In this point it is interesting to note that the solution $\Psi=\Psi_0/a^3$ implies that the fermion field has energy density $\rho_\psi=m\Psi_0/a^3$. So the extra matter field is naturally added with the common matter field. This first result corroborates to our identification of the fermion field with the dark matter.

Having in view that the Dirac's equations could be solved independently of the others field equations, in terms of $a$, one has effectively a reduction of the system (\ref{fe})-(\ref{dirac1,2}) to differential equations involving only the dynamical variables $a$ and $\phi$. Then we will solve the problem through the reduced system

\begin{equation}
3\left(\frac{\dot a}{a}\right)^2=\frac{\rho_M^0+m\Psi_0}{a^3}+\frac{\dot\phi^2}{2}+U_0\left(Ae^{\alpha\phi}-Be^{-\alpha\phi}\right)^2,
\end{equation}

\begin{equation}
\ddot\phi+3\left(\frac{\dot a}{a}\right)\dot\phi+2\alpha U_0\left(A^2e^{2\alpha\phi}-B^2e^{-2\alpha\phi}\right)=0,
\end{equation}

\begin{equation}
\left(Ae^{\alpha\phi}+Be^{-\alpha\phi}\right)\sqrt{6a}\dot a+\left(Ae^{\alpha\phi}-Be^{-\alpha\phi}\right)a^{3/2}\dot\phi=\overline{M}_0,
\end{equation}
where $\overline{M}_0=M_0+\beta_0\Psi_0$ and the result $\Psi=\Psi_0/a^3$ was used.

Hence it is enough to find a new set of variables related only to $a$ and $\phi$. An adequate particular solution of this type for the system (\ref{v1})-(\ref{v4}) is

\begin{equation}
u=a^{3/2}\frac{Ae^{\alpha\phi}-Be^{-\alpha\phi}}{\sqrt{6}AB}, \qquad v_j=0, \qquad w_j=0, \label{change1}
\end{equation}

\begin{equation}
z=a^{3/2}\frac{Ae^{\alpha\phi}+Be^{-\alpha\phi}}{\sqrt{6}AB}. \label{change2}
\end{equation}

Using (\ref{change2}) we can express the constant of motion in terms of the cyclic variable, obtaining

\begin{equation}
\dot z = \frac{\overline{M}_0}{4AB}. \label{cmc}
\end{equation}
This equation is immediately integrated, furnishing

\begin{equation}
z(t) = z_1t+z_2, \label{sol1}
\end{equation}
with $z_1=\overline{M}_0/4AB$ and $z_2=$ constant.

The Friedmann's equation takes the following form in the new variables

\begin{equation}
\dot z^2 = \dot u^2 + 3ABU_0u^2 + \frac{\rho_0+m\Psi_0}{2AB}.
\end{equation}
Substituting $\dot z$ from equation (\ref{cmc}) into the Friedmann's equation, we get its integration and obtain the solution

\begin{equation}
u(t)=u_0\sin{(\omega t+b_0)},
\end{equation}
where

\begin{equation}
u_0^2=\frac{\overline{M}_0^2-8AB\left(\rho_M^0+m\Psi_0\right)}{48A^3B^3U_0}, \qquad \omega^2=3ABU_0,\label{constants}
\end{equation}
and $b_0$ is a constant.

For the solutions $z(t)$ and $u(t)$, expressed in terms of the original variables through the relations $(\ref{change1})_1$ and (\ref{change2}), we easily get the explicit forms of $a(t)$ and $\phi(t)$

\begin{equation}
a(t)=\left(\frac{\omega^2}{2U_0}\right)^{1/3}\left\{z_1^2t^2+2z_1z_2t+z_2^2-u_0^2\sin^2{\left(\omega t+b_0\right)}\right\}^{1/3}, \label{solution1}
\end{equation}

\begin{equation}
\phi(t)=\frac{1}{2\alpha}\ln{\left\{\frac{z_1t+z_2+u_0\sin{\left(\omega t+b_0\right)}}{z_1t+z_2-u_0\sin{\left(\omega t+b_0\right)}}\right\}}-\frac{1}{2\alpha}\ln{\left(\frac{A}{B}\right)}.\label{solution2}
\end{equation}

Finally, from equation (\ref{spinor1}) one can write the time evolution of the spinor field

\begin{equation}
 \psi(t)=\frac{\sqrt{2U_0}}{\omega}\left(%
\begin{array}{c}
\psi_{1}^0e^{-\imath mt}  \\
\psi_{2}^0e^{-\imath mt}  \\
\psi_{3}^0e^{\imath mt}  \\
\psi_{4}^0e^{\imath mt}  \\
\end{array}%
\right)\Theta(t),\label{solution3}
\end{equation}
where

\begin{equation}
\Theta(t)=\left\{z_1^2t^2+2z_1z_2t+z_2^2-u_0^2\sin^2{\left(\omega t+b_0\right)}\right\}^{-1/2}.
\end{equation}

\section{Constraining the solution}

Now let us analyze the behavior of $a$. The solution (\ref{solution1}) is valid from the time when the regime is matter-dominated. Close to this time we have that the time evolution of $a$ is near the form $a\propto t^{2/3}$, i.e., when the oscillatory term $u_0^2\sin^2{(\omega t+b_0)}$ presents its minimum value and the term $z_1t$ dominates $z_2$ in the expression (\ref{solution1}). We can satisfy this limit situation by choosing the constant of integration $z_2 = 0$ and replacing the constant $b_0$ for $-\omega\tau$, such that $u_0^2\sin^2{[\omega(t-\tau)]}$ has its minimum value (zero) at $t=\tau$, where $\tau$ is the time when one has a matter-dominated regime. Before performing the analysis, one introduces the dimensionless variables: $\widetilde{t}=H_0t$, $\widetilde{\tau}=H_0\tau$, $\widetilde{m}=\frac{m}{H_0}$, $\widetilde{A}=\frac{A}{H_0}$, $\widetilde{B}=\frac{B}{H_0}$, $\widetilde{\omega}^2=3\widetilde{A}\widetilde{B}{U_0}$, $\widetilde{u}_0=H_0u_0$, $\widetilde{M_0}=\frac{\overline{M}_0}{H_0^2}$, $\widetilde{z}_2=H_0z_2$ and $\Omega_{M\psi}^0=\frac{\rho_M^0}{3H_0^2}+\frac{m\Psi_0}{3H_0^2}=\Omega_M^0+\Omega_{\psi}^0$, where $H_0$ is the Hubble parameter at the present time. Hence the solutions (\ref{solution1}), (\ref{solution2}) and (\ref{solution3}) take the forms

\begin{equation}
a(\widetilde{t}\ )=\left(\frac{\widetilde{\omega}^2}{2U_0}\right)^{1/3}\left\{z_1^2\widetilde{t}^2-\widetilde{u}_0^2
\sin^2{[\widetilde{\omega}(\widetilde{t}-\widetilde{\tau})]}\right\}^{1/3}, \label{solution1.1}
\end{equation}

\begin{equation}
\phi(\widetilde{t}\ )=\frac{1}{2\alpha}\ln{\left\{\frac{z_1\widetilde{t}+\widetilde{u}_0
\sin{[\widetilde{\omega}(\widetilde{t}-\widetilde{\tau})]}}{z_1\widetilde{t}-
\widetilde{u}_0\sin{[\widetilde{\omega}(\widetilde{t}-\widetilde{\tau})]}}\right\}}
-\frac{1}{2\alpha}\ln{\left(\frac{\widetilde{A}}{\widetilde{B}}\right)},\label{solution2.1}
\end{equation}

\begin{equation}
 \psi(\widetilde{t}\ )=\frac{\sqrt{2U_0}}{\widetilde{\omega}}\left(%
\begin{array}{c}
\psi_{1}^0e^{-\imath \widetilde{m}\widetilde{t}}  \\
\psi_{2}^0e^{-\imath \widetilde{m}\widetilde{t}}  \\
\psi_{3}^0e^{\imath \widetilde{m}\widetilde{t}}  \\
\psi_{4}^0e^{\imath \widetilde{m}\widetilde{t}}  \\
\end{array}%
\right)\left\{z_1^2\widetilde{t}^2-\widetilde{u}_0^2
\sin^2{[\widetilde{\omega}(\widetilde{t}-\widetilde{\tau})]}\right\}^{-1/2}.\label{solution3.1}
\end{equation}

From (\ref{solution1.1}), the normalization $a(\widetilde{t}_0)=1$ implies the relation

\begin{equation}
z_1^2=\frac{2U_0+\widetilde{u}_0^2\widetilde{\omega}^2
\sin{[\widetilde{\omega}(\widetilde{t}_0-\widetilde{\tau})]}}{\widetilde{\omega}^2\widetilde{t}_0^2}, \label{constr1}
\end{equation}
with $\widetilde{t}_0$ being the present time. By substituting $z_1=\widetilde{M_0}/4\widetilde{A}\widetilde{B}$ into (\ref{constr1}) and using $(\ref{constants})_1$ in the dimensionless form, written as
\begin{equation}
\widetilde{u}_0^2= \frac{3U_0\left(3U_0\widetilde{M}_0^2 - 24\widetilde{\omega}^2\Omega_{M\psi}^0\right)}{16\widetilde{\omega}^6},\label{adim_const}
\end{equation}
one obtains $\widetilde{M_0}$ in terms of $\widetilde{A}, \widetilde{B}$ and $U_0$, namely,
\begin{equation}
\widetilde{M_0}^2=\frac{8\widetilde{\omega}^2\left\{4\widetilde{\omega}^2-9\Omega_{M\psi}^0
\sin^2{[\widetilde{\omega}(\widetilde{t}_0-\widetilde{\tau})]}\right\}}{9U_0\left\{\widetilde{\omega}^2\widetilde{t}_0^2-
\sin^2{[\widetilde{\omega}(\widetilde{t}_0-\widetilde{\tau})]}\right\}}.\label{adim_const_1}
\end{equation}

  Thus we have three free parameters $\widetilde{A}, \widetilde{B}$ and $U_0$ that determine all the other parameters $\widetilde{\omega}$, $\widetilde{M}_0$, $\widetilde{u}_0$ and $z_1$ since one knows $\Omega_{M\psi}^0, \widetilde{t}_0$ and $\widetilde{\tau}$. Using the solution (\ref{solution2.1}) to rewrite the energy density and pressure of the field $\phi$, we have the following dimensionless expressions for the energy densities and pressure
\begin{eqnarray}
\widetilde{\rho}_M(\widetilde{t}\ )=&\frac{\rho_M^0}{H_0^2a(\widetilde{t}\ )^3}=\frac{3\Omega^0_M}{a(\widetilde{t}\ )^3},\qquad \widetilde{\rho}_\psi(\widetilde{t}\ )=\frac{{m}\Psi_0}{H_0^2a(\widetilde{t}\ )^3}=\frac{3\Omega^0_\psi}{a(\widetilde{t}\ )^3}
\label{adim_ed_1},\\
\widetilde{\rho}_\phi(\widetilde{t}\ )=&\frac{\widetilde{\omega}^4z_1^2\widetilde{u}_0^2}{4\alpha^2U_0^2a(\widetilde{t}\ )^6}
\left\{\sin{[\widetilde{\omega}(\widetilde{t}-\widetilde{\tau})]}
-\widetilde{\omega}\widetilde{t}\cos{[\widetilde{\omega}(\widetilde{t}-\widetilde{\tau})]}\right\}^2\nonumber\\
&+\frac{2\widetilde{\omega}^4\widetilde{u}_0^2
\sin^2{[\widetilde{\omega}(\widetilde{t}-\widetilde{\tau})]}}{3U_0a(\widetilde{t}\ )^3}\label{adim_ed_2},\\
\widetilde{p}_\phi(\widetilde{t}\ )=&\widetilde{\rho}_\phi(\widetilde{t}\ )-\frac{4\widetilde{\omega}^4\widetilde{u}_0^2
\sin^2{[\widetilde{\omega}(\widetilde{t}-\widetilde{\tau})]}}{3U_0a(\widetilde{t}\ )^3}
\label{adim_press}.
\end{eqnarray}

If we observe the solution  (\ref{solution1.1}) and the set of equations (\ref{adim_ed_1})-(\ref{adim_press}) and take into account (\ref{constr1})-(\ref{adim_const_1}), we conclude that the time evolution of the quantities (scale factor, energy densities and pressure) that will furnish the cosmological parameters depend essentially on $U_0$ and the product $\widetilde{A}\widetilde{B}$. Once the Hubble time is defined as $t_0=1/H_0$ and our dimensionless time is $\widetilde{t}=H_0t$, the approximated value for the dimensionless present time is $\widetilde{t}_0 = 1$. Note that the dimensionless present time is not exactly equal to the unity due to the fact that the Universe presents a non-constant rate of expansion (decelerated expansion when $\widetilde{t}<1$ and  accelerated expansion when $\widetilde{t}>1$). By taking the Hubble time as about 14 Gyr (see e.g. \cite{2.b,19}) and the beginning of the matter-dominated era when the Universe is about 75 kyr old, $\widetilde{\tau}$ takes the value $\widetilde{\tau}=5.4\times10^{-6}\widetilde{t}_0=5.4\times10^{-6}$. The oscillating term of $a$ has its period determined by $\widetilde{\omega}$, which can be constrained by adopting $\Omega_{M}^0=0.04$, $\Omega_{\psi}^0=0.22$ and $\Omega_{\phi}^0=0.74$. These values are based on the references \cite{2.b,21} such that at $\widetilde{t}_0$ we observe $\Omega_0=0.04+0.22+0.74=1$ in our analysis. So, from these cosmological parameters valued at the present time, one can infer the values for the free parameters. In order to present the results for a realistic description -- such that the observational data are respected --, the best values that we have found are $U_0=15$ and   $\widetilde{A}\widetilde{B}=10^{-3}$.

      \begin{figure}
 \begin{center}
 \vskip0.5cm
 \includegraphics[height=4.6cm,width=6.6cm]{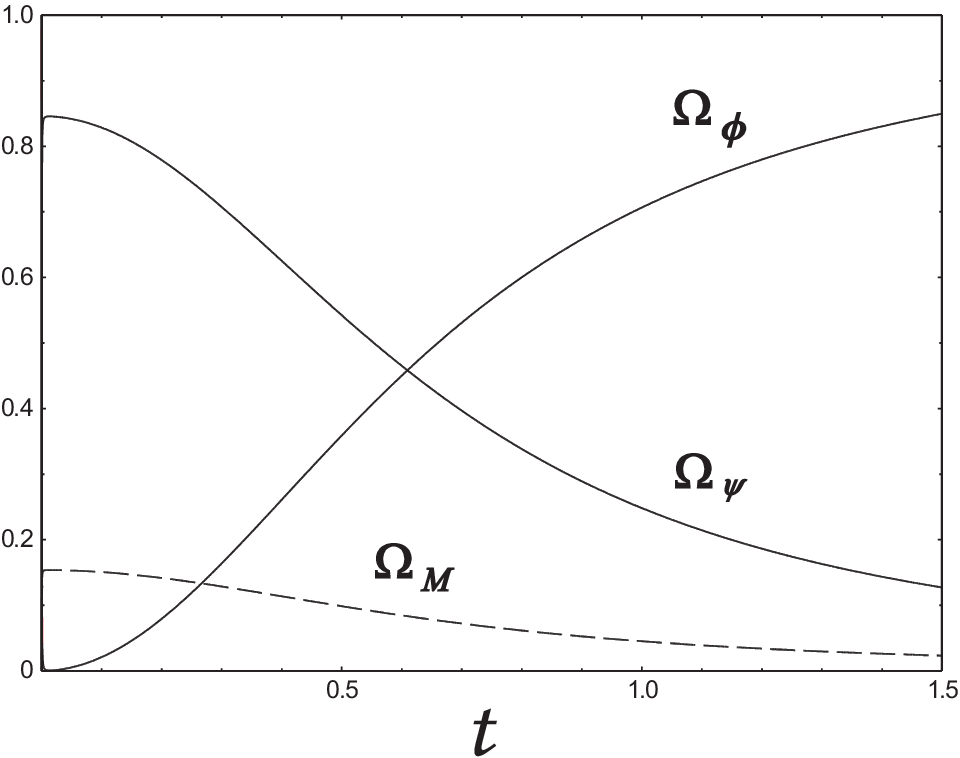}\hskip0.5cm
 \includegraphics[height=4.6cm,width=6.6cm]{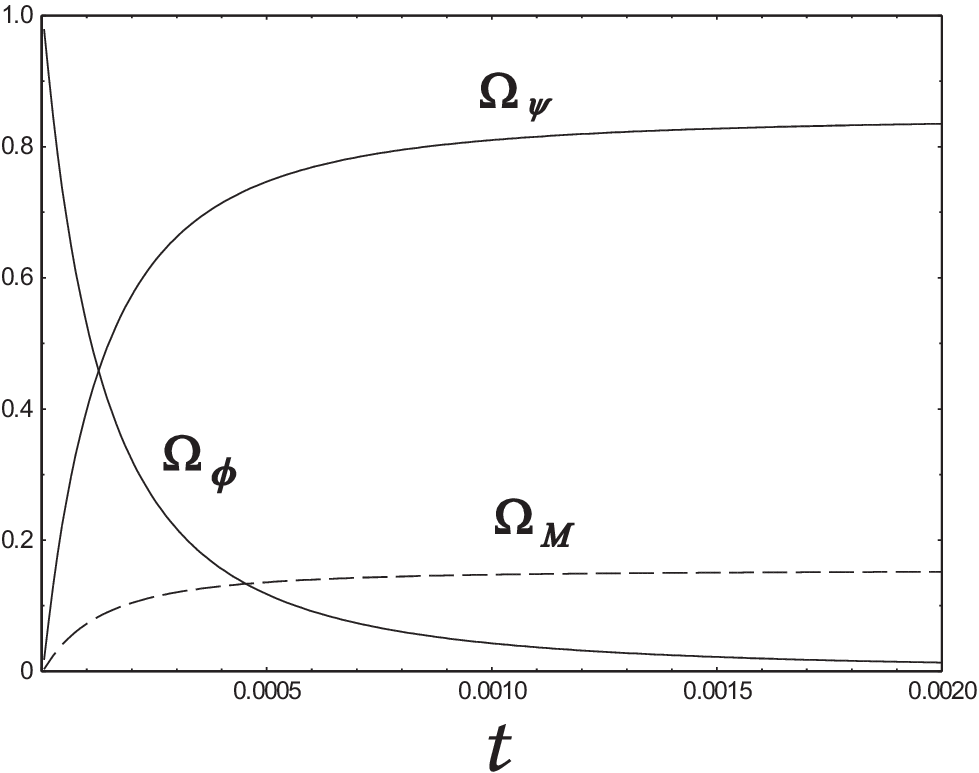}
  \caption{Density parameters of the matter fields (common matter and fermion field) and
  the boson field. Up to $t= 1.5$ (left frame) and $t=0.002$ (right frame).}
 \end{center}
 \end{figure}

  In Fig. 1 are plotted the density parameters of the matter fields (comprehending the common matter and the fermion field) and the boson field as functions of $\widetilde{t}$ -- which is just denoted by $t$ from this point on. All possible values for $U_0$ and $\widetilde{A}\widetilde{B}$ imply that $\Omega_{M}^0, \Omega_{\psi}^0$ and $\Omega_{\phi}^0$ are reached at the times larger than $t=1$. This shows that the solution naturally produces an accelerated Universe independently on the values of $U_0$ and $\widetilde{A}\widetilde{B}$. Heaving in view that the age of the Universe is close to the Hubble time, the density parameters in Fig 1 are plotted in such a form that $\Omega_{M}^0=0.04$, $\Omega_{\psi}^0=0.22$ and $\Omega_{\phi}^0=0.74$ appear as close to $t=1$ as it is possible. In Fig. 1 the present values of the density parameters are at $t=1.08$ ($\approx$ 15 Gyr), which is 8 percent larger than the Hubble time. So in the following graphics the present time will be $t=1.08$, characterizing a Universe with accelerated expansion. In the left of the Fig. 1 one can observe that $\Omega_\phi$ is dominant today, but going to the past the matter fields become the dominant components. In the right of this figure we note that close to $\tau$ one has that $\Omega_\phi$ dominates the matter density parameters. However, from the solution of $a$ we have that closing to $\tau$ the approximate behavior is $a\propto t^{2/3}$, i.e., the Universe is matter-dominated. Then the boson field must play the role of a matter field at this time. This will be clarified by the next results.

  The density parameters up to the distant future of the Universe are plotted in Fig. 2. In the left of this figure the evolution of $\Omega_\phi$ is presented, which oscillates tending to a constant value for $t\rightarrow\infty$. In the right frame the same type of evolution can be observed for $\Omega_{M\psi}$, obviously. Let us observe that in the future $\Omega_\phi$ will always dominate $\Omega_{M\psi}$.

\begin{figure}
 \begin{center}
 \vskip0.5cm
 \includegraphics[height=4.6cm,width=6.6cm]{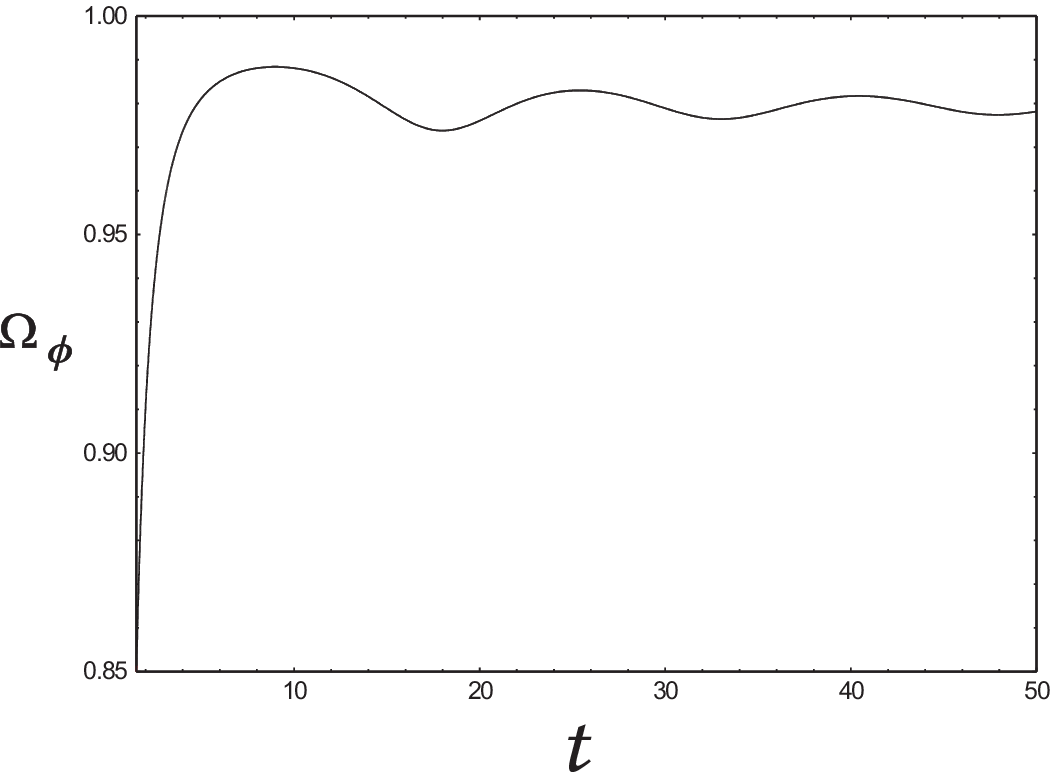}\hskip0.5cm
 \includegraphics[height=4.6cm,width=6.6cm]{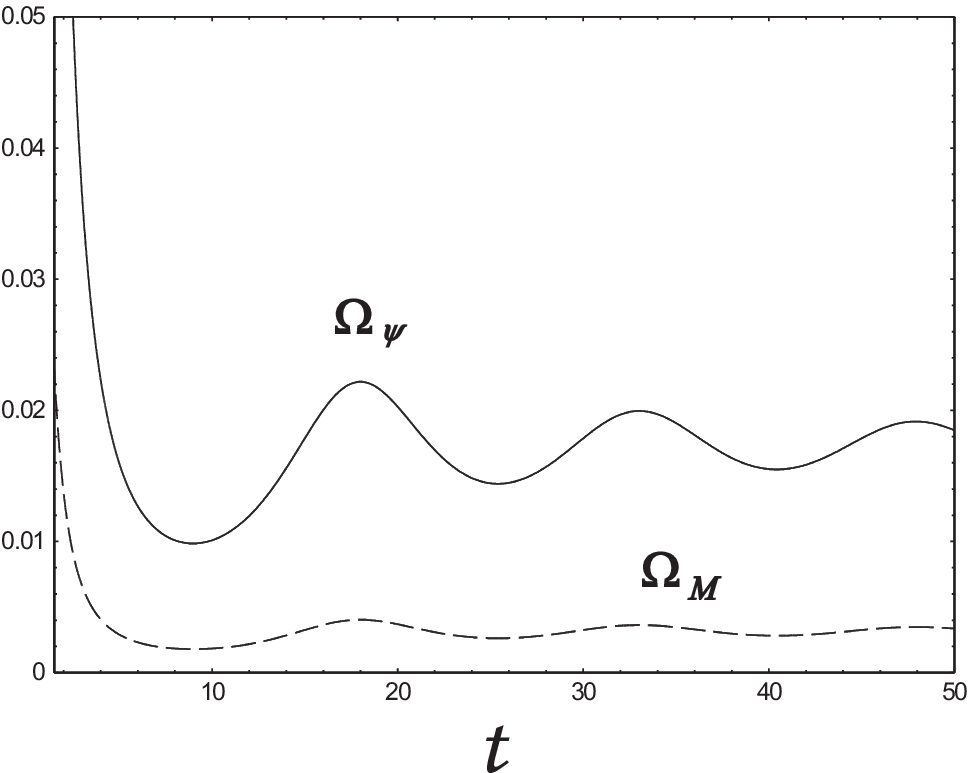}
  \caption{Density parameters of the boson field (left frame) and
  matter fields (right frame) for $t>1.5$.}
 \end{center}
 \end{figure}

The time evolution of the equation of state of the boson field $\omega_\phi = p_\phi/\rho_\phi$ is plotted in Fig. 3. From the left frame we infer that $\omega_\phi$ evolves from a positive value and decreases to $-1$ and turns to increase slowly by reaching the value $\omega_\phi=-0.73$ at the present time -- it is larger than that of the cosmological constant equation of state, $\omega_\Lambda=-1$. Note that, according
to \cite{2.b}, the value of the parameter $\omega_\phi$ inferred from the observational data for the dark energy has still great uncertainties. For times close to the beginning of the matter
dominated era, one observes from this figure that going back to $\tau$ the parameter $\omega_\phi$ reverses its sign and tends to $1$. Thus it presents a behavior of matter, explaining the approximated form $a\propto t^{2/3}$ for the scale factor with a dominant $\Omega_\phi$ in Fig. 1. In the right of this figure the time evolution of $\omega_\phi$ is presented for large times. From it one can infer that $\omega_\phi$ oscillates with time in the range $-1\leq\omega_\phi\leq1$, which shows that the boson field behaves as matter and dark energy alternately.

  \begin{figure}
 \begin{center}
 \vskip0.5cm
 \includegraphics[height=4.6cm,width=6.6cm]{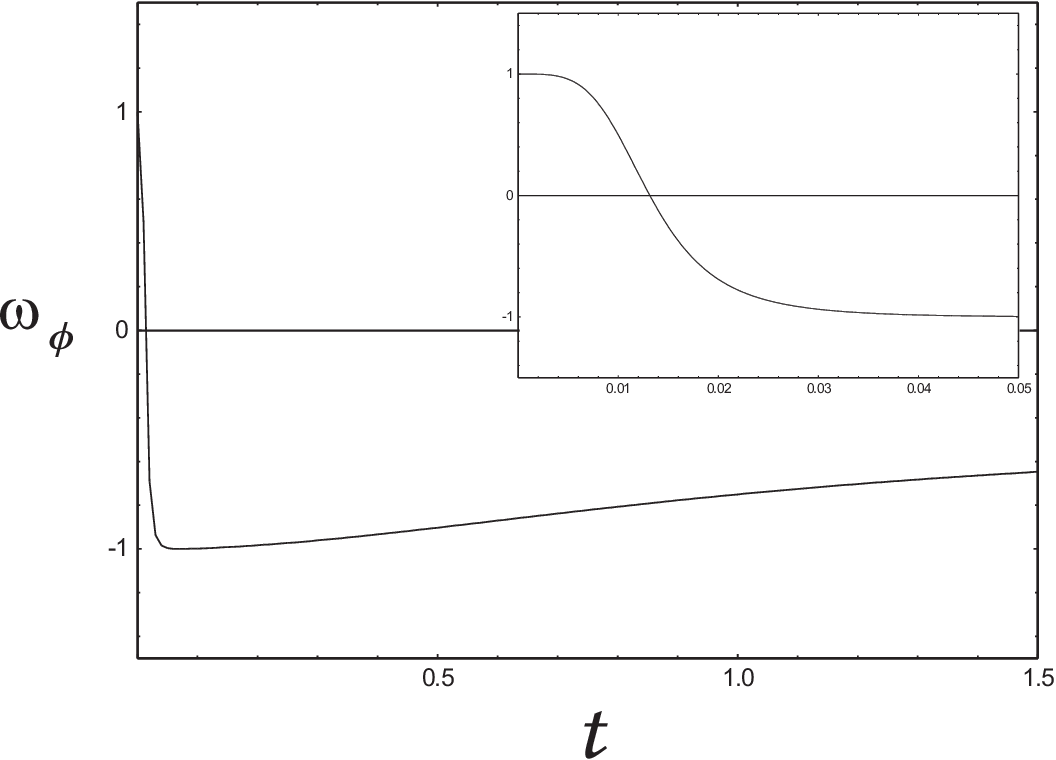}\hskip0.5cm
 \includegraphics[height=4.6cm,width=6.6cm]{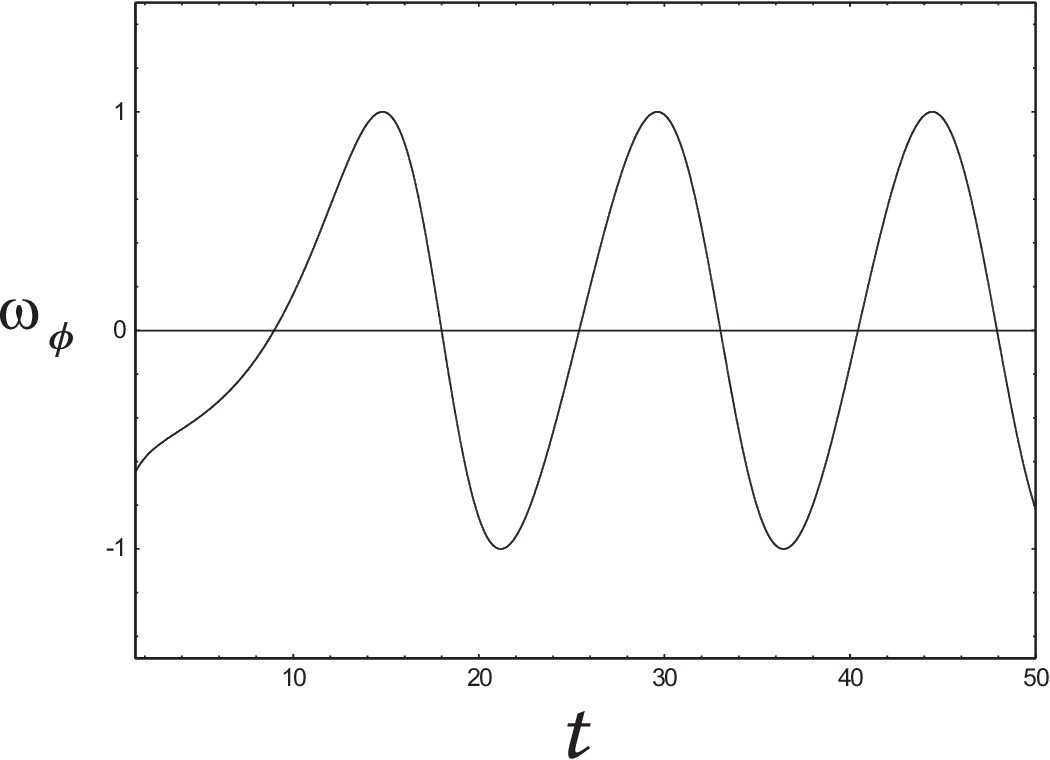}
  \caption{Equation of state of the boson field. Up to $t=1.5$ and $t=0.002$ (left frame), and for $t>1.5$ (right frame).}
 \end{center}
 \end{figure}

In Fig. 4 the acceleration of the scale factor is plotted. We started from the requirement that $\Omega_{M}^0, \Omega_{\psi}^0$ and $\Omega_{\phi}^0$ must be observed near $t=1$, being found precisely at $t=1.08$, as it was allowed by the solution. Now, in the left of Fig 4, one can see that we naturally have an accelerated expansion that have passed through a transition from a decelerated to an accelerated expansion. This transition occurs about $t=0.51$ (7.1 Gyr) and the present period is accelerated. Going back to $\tau$, an decelerated period corresponding to a real matter dominated Universe occurs. The acceleration of the scale factor for the distant future is presented in the right frame. This curve shows a Universe that will return to a decelerated period and turn to accelerate in the future and so on -- this behavior is obviously caused by the oscillating equation of state of the boson field. The final result is that the solution describes a Universe that evolves with alternated periods of accelerated and decelerated expansion and this oscillatory behavior is a kind of deadened oscillation: when $t\rightarrow\infty$, one has $\ddot a\rightarrow0$.

\begin{figure}
 \begin{center}
 \vskip0.5cm
 \includegraphics[height=4.6cm,width=6.6cm]{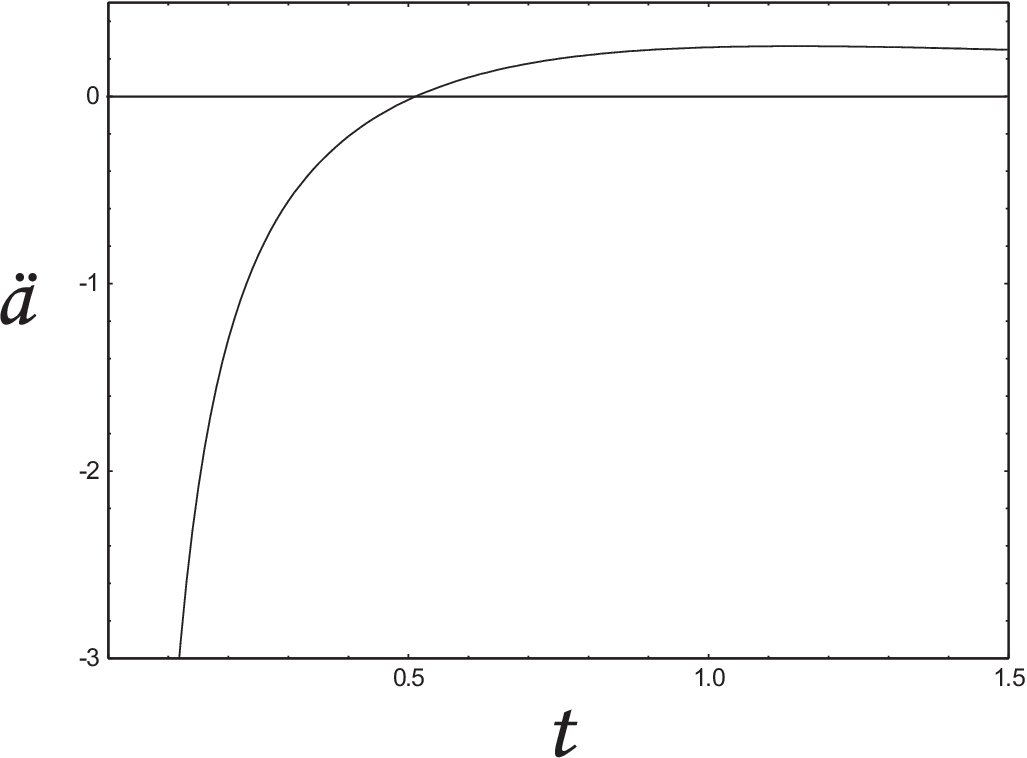}\hskip0.5cm
 \includegraphics[height=4.6cm,width=6.6cm]{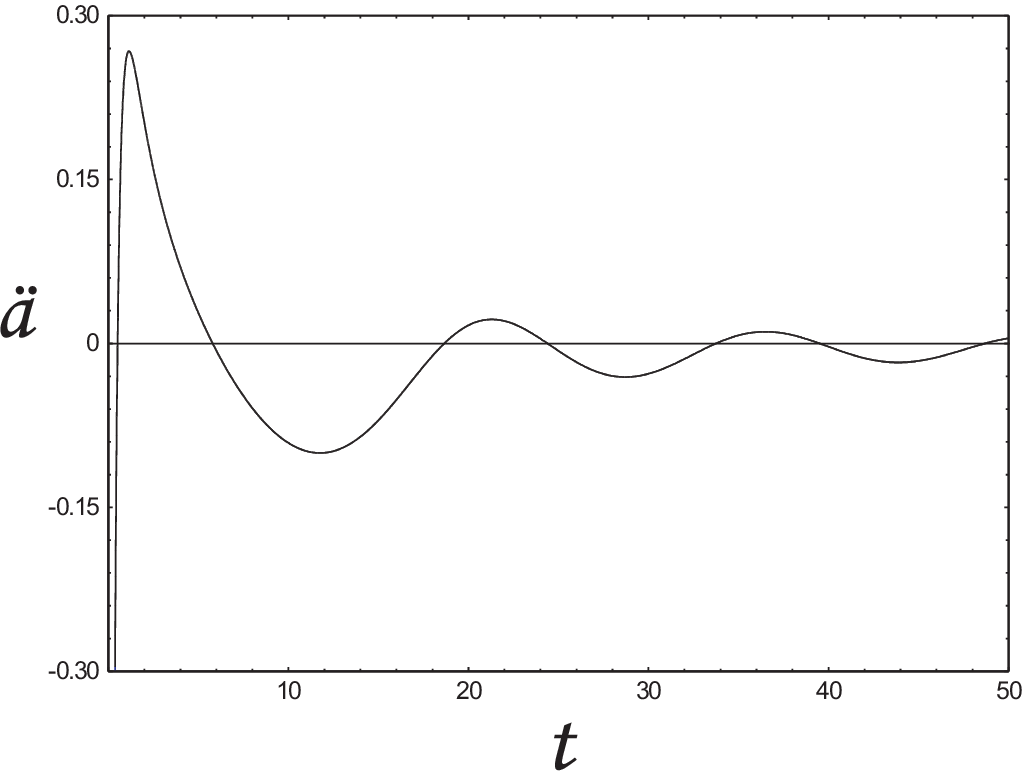}
  \caption{Acceleration of the scale factor. Up to $t=1.5$ (left frame) and
  $t=50$ (right frame).}
 \end{center}
 \end{figure}

 Now we can ask if going back in time the big bang nucleosynthesis is permissible, that is, if a radiation-dominated era can be obtained from this model. The solutions analyzed above showed that the field $\phi$ plays the role of the dominant matter field in the matter-dominated era, with the equation of state of the form $p_\phi=\rho_\phi$, informing us that at this time the referred field is kinetic dominant. To verify if a radiation era is possible from these results, we must solve the field equations for the boson and radiation fields. From what we know, such a system does not present an analytical solution for the potential $U$  used here, then we will search for a numerical solution of this scenario. Firstly, we transform the derivatives with respect to time in the Friedmann, acceleration and Klein-Gordon equations into derivatives with respect to red-shift by using the relationships
 \begin{equation}
z=\frac{1}{a}-1,\qquad \frac{d}{dt}=-H(1+z)\frac{d}{dz},
\end{equation}
and divide all of them by $H_0^2$. From this, through a little algebraic manipulation, we obtain the following system of coupled differential equations to solve
\begin{eqnarray}
2\widetilde{H}\widetilde{H}'(1+z)=\widetilde{\rho}_{R}+\widetilde{\rho}_{\phi}+\widetilde{p}_{R}+\widetilde{p}_{\phi}
,\label{ae1}\\
\widetilde{H}^2(1+z)^2\phi''+\widetilde{H}(1+z)\left[\widetilde{H}'(1+z)-2\widetilde{H}\right]\phi'+
\frac{d\widetilde{U}}{d\phi}=0,\label{phie1}
\end{eqnarray}
with $\widetilde{H}=H/H_0$, $\widetilde{U}=U/H_0^2$ and the prime denoting derivative with respect to $z$, where
\begin{eqnarray}
\widetilde{\rho}_{R}=\frac{\rho_R^0}{H_0^2a^4}=3\Omega_R^0(1+z)^4,\qquad \widetilde{p}_R=\frac{\widetilde{\rho}_{R}}{3},\\
\widetilde{\rho}_{\phi}=\frac{\widetilde{H}^2(1+z)^2\phi'^2}{2}+
U_0\left(\widetilde{A}e^{\alpha\phi}-\widetilde{B}e^{-\alpha\phi}\right)^2,\\
\widetilde{p}_{\phi}=\widetilde{\rho}_{\phi}-2U_0\left(\widetilde{A}e^{\alpha\phi}-\widetilde{B}e^{-\alpha\phi}\right)^2.
\label{pressure}
\end{eqnarray}

 From (\ref{pressure}), putting the condition $\widetilde{p}_{\phi}=\widetilde{\rho}_{\phi}$ (see Fig. 3) at $z=0$, we arrive at the condition $\phi(0)=0$ and the relation $\widetilde{A}=\widetilde{B}$. At $z=0$ we also have that $\widetilde{\rho}_{\phi}(0)=3\Omega_\phi^0=3(1-\Omega_R^0)$, which implies $\phi'(0)=\sqrt{6(1-\Omega_R^0)}$. The Friedmann equation for $z=0$ furnishes $\widetilde{H}(0)^2=\Omega_R^0+\Omega_\phi^0=1$. These are the initial conditions for the system (\ref{ae1})-(\ref{phie1}). To illustrate a solution, we adopt $U_0=15, \widetilde{A}=\widetilde{B}=0.9$ and consider at $z=0$ the boson and radiation fields as the relevant fields in the Universe, with $\Omega_R^0=\Omega_\phi^0=0.5$. Note that $z=0$ is not the present time, but the time when the important fields are the boson and the radiation.

\begin{figure}
 \begin{center}
 \vskip0.5cm
 \includegraphics[height=4.8cm,width=6.6cm]{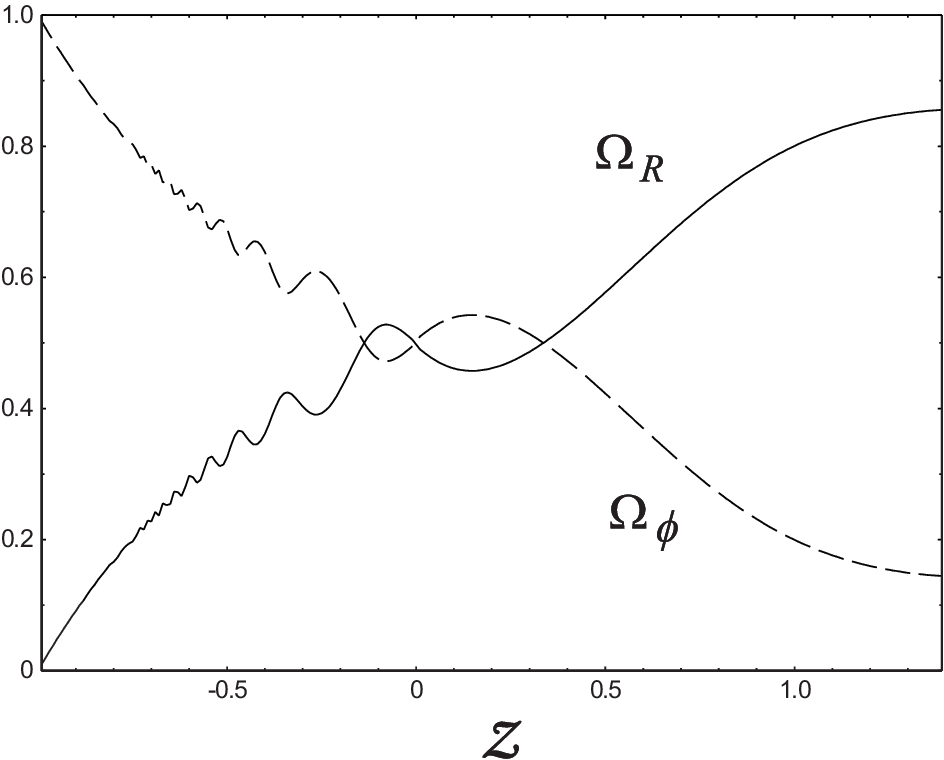}
  \caption{Density parameters of the boson (dashed line) and the radiation (straight line) fields as functions of $z$.}
 \end{center}
 \end{figure}

 In Figure 5 the density parameters of the boson and radiation fields are plotted. From this figure we can observe that with an equation of state of the form $\widetilde{p}_{\phi}=\widetilde{\rho}_{\phi}$ (kinetic dominant) for the boson field in the presence of the radiation field, if we go back in time the radiation can dominate the field $\phi$. Then the present model produces the conditions that could favor a big bang nucleosynthesis.

\section{Conclusions}

Starting from an action that describes a boson, a fermion and a common matter field, without specifying their potentials, the Noether condition is applied to the corresponding Lagrangian and the undefined potentials are constrained. With the Lagrangian satisfying the Noether symmetry, the fermion field presents a term of mass and the boson field a potential that can drive an accelerated expansion.

The first result is that the fermion field has a pressureless matter behavior and can play the role of dark matter. Additionally, the existence of a Noether symmetry proved to be useful and the complete integration of the field equations was possible. The obtained solution is then constrained by observational and physical requirements. As a result, the boson field presents an equation of state with its value varying quasi-periodically in the range $[-1,1]$, which means that it behaves as matter as well as dark energy. In the beginning of the matter dominated era the boson field plays an important role, when it behaves as a matter field. A general consequence of this solution is that the model can describe the present accelerated period and predicts a Universe with oscillating expansion. Further, the Universe evolves in such a way that in the distant future its expansion tends to a constant rate. This is, an eternal accelerated expansion as usual does not occur in this model.


\section*{References}

\begin{thebibliography}{99}

\bibitem{1}

Freese K 2000 \emph{Phys. Rep.} \textbf{333} 183


\bibitem{2}
Zwicky F 1933 \emph{Helvetica Physica Acta.} \textbf{6} 110


\bibitem{2.1}
Bergström L 2009 \emph{New J. Phys.} \textbf{11} 105006


\bibitem{2.2}
Milgrom M 2010 arXiv:1101.5122


\bibitem{2.3}
Boehmer C G, Tiberiu H and Lobo F S N 2008 \emph{Astrop. Phys.} \textbf{29} 386


\bibitem{2.4}
Bertolami O and Paramos J 2010 arXiv:1003.1875


\bibitem{3}
Clowe D \emph{et al.} 2006 \emph{Astrophys. J.} \textbf{648} L109


\bibitem{4}
Massey R \emph{et al.} 2007 \emph{Nature} \textbf{445} 286


\bibitem{5}
Riess A G \emph{et al.} 1998 \emph{Astron. J.} \textbf{116} 1009.


\bibitem{6}
Perlmutter S \emph{et al.} 1999 \emph{Astrophys. J.} \textbf{517} 565


\bibitem{2.a}

Bertolami O 2009 \emph{Int. J. Mod. Phys.} \textbf{18} 2303.


\bibitem{2.b}


Lahav O and Liddle A R 2010 arXiv:1002.3488


\bibitem{7}
Peebles P J E and Ratra B 2003 \emph{Rev. Mod. Phys.} \textbf{75} 559


\bibitem{8}
Szydlowski M Kurek A and Krawiec A 2006 \emph{Phys. Lett. B} \textbf{642} 171


\bibitem{8.1}
Brax P Martin J and Riazuelo A 2000 \emph{Phys Rev. D} \textbf{62} 103505


\bibitem{8.2}
Gardner C L 2005 \emph{Nucl. Phys. B} \textbf{707} 278


\bibitem{8.3}
Henttunen K Multamäki T and Vilja I 2006 \emph{Phys. Lett. B} \textbf{634} 5


\bibitem{7.3}

Ribas M O Devecchi F P and Kremer G M 2005 \emph{Phys. Rev. D} {\textbf{72}} 123502


\bibitem{7.3.1}

Saha B 2006 \emph{Grav. Cosmol.} {\textbf{12}} {215}


\bibitem{7.3.2}

Ribas M O Devecchi F P and Kremer G M 2008 \emph{Europhys. Lett.} {\textbf{81}} {19001}


\bibitem{7.4}

Ren J and Meng X H 2008 \emph{Int. J. Mod. Phys. D} {\textbf{17}} 2325


\bibitem{7.4.1}

de Souza R C and Kremer G M 2009 \emph{Class. Quant. Grav.} {\textbf{26}} {135008}


\bibitem{7.6}

Avelino P P Beca L M G de Carvalho J P M and Martins C J A P 2003 \emph{JCAP} {\textbf{0309}} {002}


\bibitem{7.6.1}

Kremer G M 2003 \emph{Gen. Rel. Grav.} {\textbf{35}} {1459}


\bibitem{7.a}

Capozziello S de Martino S and Falanga M 2002 \emph{Phys. Lett. A} {\textbf{299}} {494}


\bibitem{7.b}

Kremer G M 2003 \emph{Phys. Rev. D} {\textbf{68}} {123507}


\bibitem{7.7}

Capozziello S and de Felice A 2008 \emph{JCAP} {\textbf{0808}} {016}


\bibitem{7.7.1}

Sotiriou T P and Faraoni V 2010 \emph{Rev. Mod. Phys.} {\textbf{82}} {451}


\bibitem{9}
Capozziello S and de Ritis R 1994 \emph{Class. Quantum Grav.} \textbf{11} 107


\bibitem{9.1}
Fy S 2001 \emph{Class. Quantum Grav.} \textbf{18} 4863


\bibitem{11}
Capozziello S Dunsby P K S Piedipalumbo E and Rubano C 2007 \emph{Astron. and
Astroph.} \textbf{472} 51


\bibitem{12}
de Souza R C and Kremer G M 2008 \emph{Class. Quantum Grav.} \textbf{25} 225006


\bibitem{10}
Zhang Y Gong Y Z and Zhu Z H 2010 \emph{Phys. Lett. B} \textbf{688} 13


\bibitem{13.1}
Vakili B 2010 \emph{Ann. Phys.} \textbf{19} 359


\bibitem{14}
Micheletti S M R Abdalla E and Wang B 2009 \emph{Phys. Rev. D} \textbf{79} 123506


\bibitem{12.1}
Armendariz-Picon C and Greene P B 2003 \emph{Gen. Rel. Grav.} \textbf{35} 1637


\bibitem{12.2}
 Saha B 2010 \emph{Roman. Rep. Phys.} \textbf{62} 209


\bibitem{19}

Coles J 2008 \emph{Astrophys. J.} \textbf{679} 17


\bibitem{21}

Liang N Wu P and Zhang S N 2010 \emph{Phys. Rev. D} \textbf{81} 083518



\end{thebibliography}
\end{document}